\documentclass[12pt,notitlepage,aip,php,amsmath,amssymb]{revtex4-1}

\usepackage{graphicx}

\usepackage{times}

\newcommand{\etal}{\textit{et al.}}

\newcommand{\Eqref}[1]{Eq.~\eqref{#1}}
\newcommand{\Eqsref}[1]{Eqs.~\eqref{#1}}
\newcommand{\Figref}[1]{Fig.~\ref{#1}}
\newcommand{\Figsref}[1]{Figs.~\ref{#1}}

\newcommand{\Sn}{\Phi} 
\newcommand{\Snw}{\varphi} 

\newcommand{\dSn}{\Theta} 
\newcommand{\dSnw}{\vartheta} 

\newcommand{\nSn}{\widehat{\Sn}}

\newcommand{\mean}[1]{\left< {#1} \right>}
\newcommand{\angs}[1]{\langle {#1} \rangle} 

\newcommand{\mA}{\left< A \right>} 
\newcommand{\mSn}{\left< \Sn \right>} 
\newcommand{\rSn}{\Sn_{\text{rms}}} 

\newcommand{\rms}{\text{rms}}
\newcommand{\rmd}{\text{d}} 
\newcommand{\td}{\tau_{\text{d}}} 
\newcommand{\tw}{\tau_{\text{w}}} 
\newcommand{\eX}{X} 
\newcommand{\avT}{\left<\Delta T\right>} 
\newcommand{\avTangs}{\langle \Delta T\rangle} 

\DeclareMathOperator\erfc{erfc}

\newcommand{\JGR}{\textit{J.~Geophys.~Res.}}
\newcommand{\JNM}{\textit{J.~Nuclear Mater.}}

\newcommand{\NF}{\textit{Nucl.\ Fusion}}

\newcommand{\PPCF}{\textit{Plasma Phys.\ Contr.\ Fusion}}

\newcommand{\PP}{\textit{Phys.\ Plasmas}}
\newcommand{\PS}{\textit{Phys.\ Scripta}}

\newcommand{\BTSJ}{\textit{Bell Sys.\ Tech. J.}}

\newcommand{\PRL}{\textit{Phys.~Rev.\ Lett.}}
\newcommand{\BLM}{\textit{Boundary Layer Meteorol.}}

\begin{document}

\title{Level crossings, excess times and transient plasma--wall interactions in fusion plasmas}



\author{A.~Theodorsen}

\author{O.~E.~Garcia}

\affiliation{Department of Physics and Technology, UiT The Arctic University of Norway, N-9037 Troms{\o}, Norway}

\date{\today}

\begin{abstract}
Based on a stochastic model for intermittent fluctuations in the boundary region of magnetically confined plasmas, an expression for the level crossing rate is derived from the joint distribution of the process and its derivative. From this the average time spent by the process above a certain threshold level is obtained. This provides novel predictions of plasma--wall interactions due to transient transport events associated with radial motion of blob-like structures in the scrape-off layer.
\end{abstract}

\maketitle
Plasma--wall interactions remain an outstanding challenge in the quest for controlled thermonuclear fusion based on magnetic confinement. \cite{pitts,lipschultz,dmz} Transient transport events due to filamentary structures moving through the scrape-off layer may cause detrimental sputtering and erosion of the main chamber walls. The interaction between the hot plasma and material surfaces depends on the turbulence-induced particle and heat fluxes, so gaining insight into the statistical properties of plasma fluctuations in the boundary region is of considerable interest.

The radial propagation of blob-like structures results in large-amplitude bursts in single-point measurements in the scrape-off layer. Recent analysis of such measurement time series using conditional averaging have elucidated the statistical properties of large-amplitude fluctuations. \cite{garcia-acm,garcia-tcv,theodorsen,kube} The experimental results provide evidence that plasma fluctuations can be described as a super-position of uncorrelated pulses with fixed, exponential pulse shape of constant duration and exponentially distributed pulse amplitudes. These are the basic assumptions behind a recently suggested stochastic model for intermittent plasma fluctuations in the scrape-off layer region. \cite{rice,garcia-shotnoise,kg} 
This model describes many experimental findings from the boundary region of magnetized plasmas, including bursty fluctuations, skewed and flattened probability density functions and accordingly a parabolic relation between the skewness and flatness moments for a broad range of parameters.\cite{garcia-acm,garcia-tcv,theodorsen,kube,sattin,bergsaker}

Based on this stochastic model, the joint distribution function of the process and its derivative is derived. This is shown to give novel predictions of the intermittent features of plasma fluctuations, in particular the rate of level crossings and excess time statistics, that is, the duration of time intervals where the signal exceeds some prescribed threshold level. \cite{rice-level,kristensen,sato,fattorini,dalmao} Although of particular interest for plasma--wall interactions in fusion grade plasmas, the stochastic model is prototypical for many intermittent systems, and the results find applications in a broad range of fields (see for example Ref.~\onlinecite{dalmao} and references therein).

%
%

Given the joint probability density function $P_{\Sn \dot{\Sn}}(\Sn,\dot{\Sn})$ for a stationary random variable $\Sn(t)$ and its derivative $\dot{\Sn}=\rmd\Sn/\rmd t$, the number of up-crossings of the level $\Phi$ in a time interval of duration $T$ is given by \cite{rice-level,kristensen,sato,fattorini}
\begin{equation}\label{rice}
\eX(\Phi) = T \int\limits_{0}^{\infty} \text{d}\dot{\Sn}\, \dot{\Sn} P_{\Sn \dot{\Sn}} ( \Sn,\dot{\Sn} ) .
\end{equation}
For independent, normally distributed $\Sn$ and $\dot{\Sn}$, this gives the celebrated result known as the Rice formula, \cite{rice-level,kristensen,sato,fattorini,dalmao}
\begin{equation} \label{ricenorm}
\eX(\Sn) = T\, \frac{\dot{\Sn}_\rms}{2\pi\Sn_\rms} \exp{\left( - \frac{\left( \Sn-\mSn \right)^2}{2 \rSn^2} \right)} ,
\end{equation}
where $\angs{\Sn}$ is the mean value of $\Sn$ and $\Sn_\rms$ and $\dot{\Sn}_\rms$ are the root mean square (rms) values of $\Sn$ and $\dot{\Sn}$, respectively. The rate of level crossings is clearly largest for threshold values close to the mean value of $\Sn$.

The average time $\avTangs$ spent above a threshold value $\Phi$ by the stationary process is given by the ratio of the total time spent above the level $\Sn$ and the number of up-crossings $\eX$ in an interval of duration $T$. The former is by definition given by the complementary cumulative distribution function $1-C_\Sn$ for the process, where $C_\Sn(\Sn)$ is the cumulative distribution function. This gives the average excess time as
\begin{equation}
\avT\left( \Sn \right) =T\, \frac{1-C_\Phi(\Phi)}{\eX\left( \Sn \right)} .
\end{equation}
For independent, normally distributed $\Sn$ and $\dot{\Sn}$, the average excess time is given by \cite{kristensen,sato,fattorini}
\begin{equation}
  \avT\left( \Sn \right) = \pi\,  \frac{\rSn}{\dot{\Sn}_\rms}\erfc\left( \frac{\Sn-\mSn}{\sqrt{2}\rSn} \right)\exp\left( \frac{\left(\Sn-\mSn\right)^2}{2 \rSn^2} \right) ,
\end{equation}
where $\text{erfc}$ denotes the complementary error function. This normal limit has previously been compared with measurement data from a basic laboratory experiment and rocket data from the polar ionosphere, and the discrepancy interpreted as a signature of intermittency in the underlying processes. \cite{sato,fattorini}

The goal of this contribution is to generalize the above expressions for level crossings and excess times for a stochastic process that describes intermittent fluctuations in the boundary region of magnetically confined plasmas. The plasma fluctuations are in this case described as a super-position of uncorrelated pulses, \cite{rice,garcia-shotnoise,kg}
\begin{equation} \label{PhiK}
\Sn_K(t) = \sum_{k=1}^{K(T)} A_k \Snw\left( t-t_k \right) ,
\end{equation}
where $t_k$ is the pulse arrival time for event $k$, $A_k$ is the pulse amplitude and the pulse shape $\Snw(t)$ is assumed to be the same for all events. In \Eqref{PhiK} the sum is over exactly $K$ pulses present in a record of duration $T$, and the pulse arrival times are assumed to have a uniform distribution. From this it follows that the number of pulses $K(T)$ is  Poisson distributed with constant rate $1/\tw$,
\begin{equation}
P_K(K) = \frac{1}{K!} \left( \frac{T}{\tw} \right)^K \exp\left( -\frac{T}{\tw} \right).
\end{equation}
Thus, the waiting time between pulses are exponentially distributed with mean value $\tw$.

In the following, the pulse shapes are described by a double-exponential function
\begin{equation}\label{pulse}
\Snw(t) = 
\begin{cases}
\exp{\left(  t / \lambda \td \right)}, & \quad t<0 ,
\\
\exp{\left( - t/(1-\lambda)\td \right)}, & \quad t \geq 0 ,
\end{cases}
\end{equation}
where $\td$ is the pulse duration and $\lambda$ is a pulse shape asymmetry parameter restricted to the range $0<\lambda<1$. The ratio between the pulse duration and average waiting time,
\begin{equation}
\gamma = \frac{\td}{\tw} ,
\end{equation}
determines the degree of pulse overlap and is the most fundamental parameter of the stochastic model.

Given exponentially distributed pulse amplitudes with mean value $\mA$, the stationary distribution of the random variable $\Phi(t)$ can be shown to be Gamma distributed with shape parameter $\gamma = \td/\tw$ and scale parameter $\mA$; \cite{garcia-shotnoise}
\begin{equation}
  P_\Sn(\Sn) = \frac{1}{\mA \Gamma(\gamma)} \left( \frac{\Sn}{\mA} \right)^{\gamma-1} \exp\left( -\frac{\Phi}{\mA} \right).
  \label{eq:pdf_shotnoise}
\end{equation}
The mean of the random variable $\Sn$ is $\mSn = \gamma \mA$ and the variance is $\Sn_\rms^2 = \gamma \mA^2$, giving the relative fluctuation level $\rSn / \mSn = 1/\gamma^{1/2}$. The skewness of $\Sn$ is $S_\Sn = 2/\gamma^{1/2}$ and the flatness is $F_\Sn = 3+6/\gamma$, giving a parabolic relation between skewness and flatness: $F_\Sn\left( S_\Sn \right) = 3+3 S_\Sn^2 / 2$.
This parabolic relation is a very good description of experimental data from the scrape-off layer.\cite{garcia-shotnoise,garcia-acm,kube,theodorsen,sattin}
It can be shown that the distribution of the normalized process $\nSn = (\Sn-\mSn)/\rSn$ resembles a normal distribution in the limit $\gamma \to \infty$, independent of pulse shape and amplitude distribution. In this case, both the skewness $S_{\Sn}$ and the excess kurtosis $F_{\Sn}-3$ vanish.\cite{rice,garcia-shotnoise}

Note that in the case of positive definite amplitudes and the pulse shape in \Eqref{pulse},  $\Sn$ is non-negative, giving $\nSn \geq - \gamma^{1/2}$. By contrast, a normally distributed random variable has infinite support. The difference between the probability density function of $\nSn$ and a standard normal distribution (the distribution of a normally distributed variable with zero mean and unit standard deviation) due to this discrepancy is negligible, however, since values of $-\gamma^{1/2}$ or less are highly unlikely for a standard normal distribution in the case of $\gamma \gg 1$.

\begin{figure}
  \centering
  \includegraphics[width = 8cm]{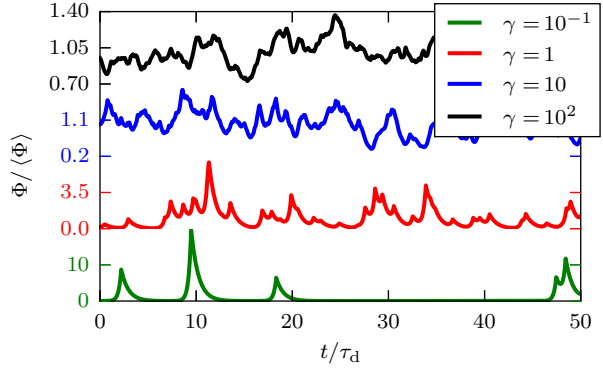}
  \caption{Realizations of the stochastic process for $\lambda = 1/4$ and various values of $\gamma$.}
  \label{fig:realizations}
\end{figure}

Realizations of this process for various values of $\gamma$ are shown in \Figref{fig:realizations}. For small $\gamma$, the pulses are well separated and the process is strongly intermittent. For large $\gamma$, there is significant pulse overlap and realizations of the process resembles random noise, with relatively small and symmetric fluctuations around the mean value. The parameter $\gamma$ can therefore be interpreted as an intermittency parameter for the process.

In order to calculate the joint distribution of the process and its derivative, the normalized time derivative is defined by
\begin{equation}
\dSn_K(t) = \td \frac{\text{d}\Sn_K}{\text{d} t} 
= \sum_{k = 1}^{K(T)} A_k \dSnw{\left( t-t_k \right)} ,
\end{equation}
where the pulse shape is given by
\begin{equation}
\dSnw(t) =
\begin{cases}
\lambda^{-1} \exp{\left(  t / \lambda\td \right)}, & \quad t<0 ,
\\
-(1-\lambda)^{-1} \exp{\left( - t/(1-\lambda)\td \right)}, & \quad t \geq 0 .
\\
\end{cases}
\end{equation}
This is another stochastic process of the same type as given in \Eqref{PhiK}, but with a different pulse shape. Since the process $\Sn(t)$ is stationary, it follows that $\mean{\dSn}=0$. The processes $\Sn(t)$ and $\dSn(t)$ are evidently dependent yet also uncorrelated,
\begin{equation}
\mean{\Sn \dSn} = \frac{\td}{2}\frac{\text{d}}{\text{d}t} \mean{\Sn^2} = 0 .
\end{equation}
The lowest order moments of $\dSn$ are readily calculated as $\dSn_\rms^2 = \gamma \mA^2 /\lambda (1-\lambda)$, $S_\dSn = 2(1-2\lambda)/[\gamma \lambda (1-\lambda)]^{1/2}$ and $F_\dSn = 3+6[1+(1-2\lambda)^2/\lambda (1-\lambda)]/\gamma$. Like before, it is possible to show that the probability density function of $\dSn$ resembles a normal distribution in the limit $\gamma \to \infty$.

%
%

Using that individual events are uncorrelated and that the number of pulses is Poisson distributed, the joint probability density function of $\Sn$ and $\dSn$ can be calculated as
\begin{equation}
  P_{\Sn \dSn} \left( \Sn, \dSn \right) = \frac{1}{\left( 2 \pi \right)^2} \int\limits_{-\infty}^{\infty} \text{d} u \, \int\limits_{-\infty}^{\infty} \text{d}v \, \exp\left( -i \Sn u -i \dSn v \right) \mean{\exp\left( i u \Sn + i v \dSn \right)},
  \label{jpdf-start}
\end{equation}
where
\begin{equation}
  \mean{\exp\left( i u \Sn + i v \dSn \right)} = \exp\left( \frac{1}{\tw} \int\limits_{-\infty}^{\infty}\text{d}A P_A(A)\, \int\limits_{-\infty}^{\infty} \text{d}t\, \left[  \exp\left( i u A \Snw(t) + i v A \dSnw(t) \right)-1 \right]  \right)
  \label{jpdf-charfun-start}
\end{equation}
is the joint characteristic function between $\Sn$ and $\dSn$. This expression is given in Ref.~\onlinecite{sato} for the case of fixed (degenerately distributed) pulse amplitudes, although the generalization is straightforward. For the process described here, a lengthy calculation gives
\begin{equation}
  \mean{\exp\left(i u \Sn +i v \dSn \right)} = \left[ 1 - i \mA \left( u + \frac{v}{\lambda} \right) \right]^{-\gamma \lambda} \left[ 1- i \mA \left( u-\frac{v}{1-\lambda} \right) \right]^{-\gamma\left( 1-\lambda \right)}.
  \label{jpdf-charfun}
\end{equation}
Substituted into \Eqref{jpdf-start}, the stationary joint probability density function can be obtained in closed form. This is non-zero only for the limited range $-\Sn/(1-\lambda)<\dSn<\Sn/\lambda$, and given by 
\begin{equation} \label{jpdf}
  P_{\Sn \dSn}{\left( \Sn, \dSn \right)} = \frac{\gamma^\gamma \lambda^{\gamma \lambda} (1-\lambda)^{\gamma(1-\lambda)}  }{\mSn^\gamma \Gamma(\gamma \lambda) \Gamma{\left( \gamma (1-\lambda) \right)}} \exp\left(-\frac{\gamma\Sn}{\mSn}\right) \left[ \Sn+(1-\lambda)\dSn \right]^{\gamma \lambda -1} \left( \Sn-\lambda \dSn \right)^{\gamma (1-\lambda) -1}.
\end{equation}
This limited range of the non-zero joint probability follows from the fact that the signal $\Sn(t)$ cannot decrease faster than the rate of decay of individual pulse structures, nor increase slower than the rate of growth of individual pulses.

As the probability density function of both $\Sn$ and $\dSn$ resembles a normal distribution in the limit $\gamma \to \infty$ and they are uncorrelated, the joint probability density function for $\Sn$ and $\dSn$ resembles the product of two normal distributions, that is, a joint normal distribution with vanishing correlation coefficient. Thus, in the normal limit $\gamma\to\infty$, the classical Rice formula given by \Eqref{ricenorm} discussed above is recovered. 
As in the case of $P_\Sn$, there is a discrepancy between $P_{\Sn \dSn}$ and a joint normal distribution due to the limited region of non-zero values of $P_{\Sn \dSn}$. The domain of non-zero values can be written as $-(\nSn+\gamma^{1/2})/(1-\lambda) < \lambda (1-\lambda) \widehat{\dSn}  < (\nSn+\gamma^{1/2})/\lambda $ where $\widehat{\dSn} = \dSn/\dSn_\rms$. For standard normally distributed variables, values outside of this domain are highly unlikely in the case of $\gamma \gg 1$, and this discrepancy is in practice negligible.

\begin{figure}
  \centering
    \includegraphics[width = 8cm]{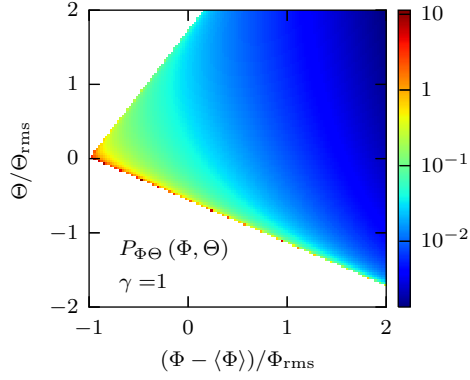}
    \caption{The joint probability density function for the stochastic process and its derivative for $\lambda=1/4$ and $\gamma=1$.}
    \label{fig:jpdf-g1}
\end{figure}

\begin{figure}
  \centering
    \includegraphics[width = 8cm]{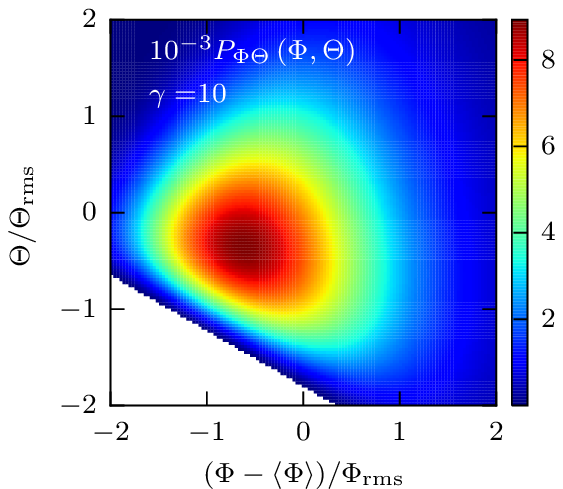}
  \caption{The joint probability density function for the stochastic process and its derivative for $\lambda=1/4$ and $\gamma=10$.}
  \label{fig:jpdf-g10}
\end{figure}

The joint distribution $P_{\Sn\dSn}(\Sn,\dSn)$ is presented in \Figsref{fig:jpdf-g1} and~\ref{fig:jpdf-g10} for $\gamma=1$ and $\gamma=10$, respectively. It should be noted that logarithmic scaling is used in \Figref{fig:jpdf-g1} while linear  scaling is used in \Figref{fig:jpdf-g10}. The white area in both figures are the regions where $P_{\Sn\dSn}$ vanishes, as given by \Eqref{jpdf}. The joint distribution for $\gamma \leq 1$ diverges at $\Sn = 0$ and $\dSn = 0$, since the pulses arrive rarely enough for the signal to fall close to zero for long time durations. In this case, the signal is very likely to decay undisturbed at the rate of individual pulses, explaining the increased value of the joint distribution near the line $\dSn = -\Sn/(1-\lambda)$. The joint distribution for $\gamma = 10$ is unimodal, since significant pulse overlap causes a wider range of values for $\dSn$ to be likely for a given value of $\Sn$.

The rate of up-crossings above a threshold $\Sn$ is now readily calculated from \Eqref{rice} as
\begin{equation}
\frac{\td}{T} \eX(\Phi) = \int_{0}^{\infty}\text{d}\dSn\, \dSn P_{\Sn \dSn}(\Phi, \dSn) = \frac{\lambda^{\gamma \lambda -1} \left( 1-\lambda \right)^{\gamma\left( 1-\lambda \right)-1}}{\gamma \Gamma\left( \gamma \lambda \right) \Gamma\left( \gamma \left( 1-\lambda \right) \right)} \left( \frac{\gamma \Phi}{\mSn} \right)^\gamma \exp\left( - \frac{\gamma\Phi}{\mSn} \right),
\label{eq:sn_X}
\end{equation}
which, together with the complementary cumulative distribution function of the Gamma distributed variable $\Sn$, 
\begin{equation}
  1-C_\Sn(\Sn) = Q\left( \gamma , \gamma \Sn/\mSn \right),
  \label{eq:sn_ccdf}
\end{equation}
where $Q$ is the regularized upper gamma function, gives the average time above the threshold
\begin{equation}
\frac{1}{\td}\,\avT(\Phi) = \frac{\gamma \Gamma\left(\gamma\lambda\right)\Gamma\left(\gamma(1-\lambda)\right)}{\lambda^{\gamma \lambda-1} (1-\lambda)^{\gamma (1-\lambda)-1}} Q\left( \gamma, \frac{\gamma \Phi}{\mSn} \right) \left( \frac{\gamma \Phi}{\mSn} \right)^{-\gamma}\exp\left( \frac{\gamma \Phi}{\mSn} \right).
\label{eq:sn_avT}
\end{equation}
Note that both \Eqsref{eq:sn_X} and \eqref{eq:sn_avT} can be written as a pre factor depending on $\gamma$ and $\lambda$ multiplied by a function of $\gamma$ and $\Sn/\mSn$. This indicates that the functional shape of both equations with threshold level depends only on the intermittency parameter $\gamma$ while the total value of the functions depends on both $\gamma$ and $\lambda$. In contrast, the complementary cumulative distribution function \Eqref{eq:sn_ccdf} does not depend on $\lambda$. Thus we only present $X$, $1-C_\Sn$ and $\avT$ for fixed $\lambda$ and various values of $\gamma$ in the following.

\begin{figure}
  \centering
    \includegraphics[width = 8cm]{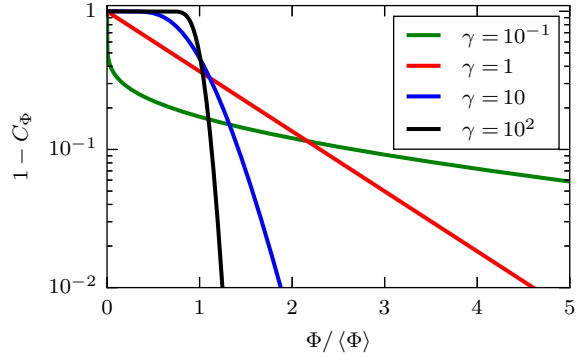}
  \caption{The complementary cumulative distribution function of the stochastic process for various values of $\gamma$.}
  \label{fig:ccdfSn}
\end{figure}

The complementary cumulative distribution function as a function of the threshold level for various values of $\gamma$ is presented in \Figref{fig:ccdfSn}. As $\gamma$ increases, this function approaches that of a normal distribution and, in the normal regime $\gamma \gg 1$, the fraction of time above threshold falls rapidly with increasing threshold level since the fluctuations in the signal are concentrated around the mean value. In the strong intermittency regime, $\gamma \ll 1$, the signal spends long periods of time close to zero value as few pulses overlap. Thus the total time above threshold increases rapidly as the threshold approaches zero.
 
\begin{figure}
  \centering
    \includegraphics[width = 8cm]{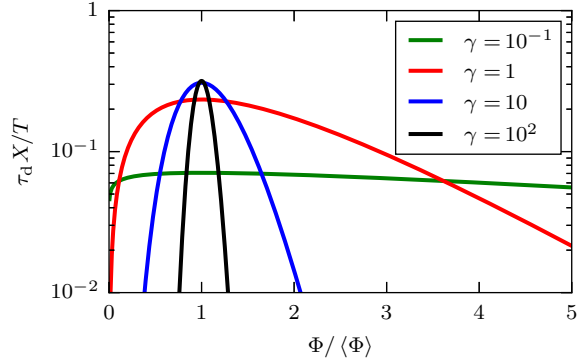}
  \caption{The rate of up-crossings for the stochastic process with $\lambda = 1/4$ and various values of $\gamma$.}
  \label{fig:ex-X}
\end{figure}

The rate of up-crossings as function of the threshold level for various values of $\gamma$ is presented in \Figref{fig:ex-X}. The number of crossings is evidently proportional to the length of the time series $T$ and inversely proportional to the pulse duration $\td$. The rate of threshold crossings is highest for thresholds close to the mean value of the process in all cases. In the non-intermittent regime $\gamma\gg1$, there are few crossings for threshold levels much smaller or much larger than the mean value due to the low probability of large-amplitude fluctuations. The rate of level crossings is therefore a narrow normal distribution in this limit. In the strong intermittency regime, $\gamma\ll1$, the signal spends most of the time close to zero value, and virtually any pulse arrival will give rise to a level crossing for finite threshold values. As seen in \Figref{fig:ex-X}, the rate of level crossings approaches a step function in this limit. 

\begin{figure}
  \centering
    \includegraphics[width = 8cm]{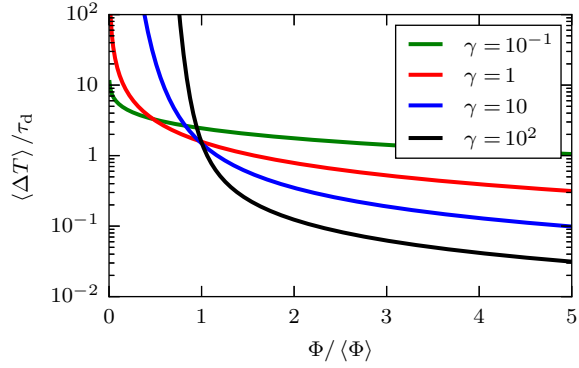}
  \caption{The average time above threshold for the stochastic process with $\lambda = 1/4$ and various values of $\gamma$.}
  \label{fig:ex-avT}
\end{figure}

The average time above threshold is presented in \Figref{fig:ex-avT} for various values of $\gamma$. While both the rate of threshold crossings and the fraction of time above threshold vary qualitatively as $\gamma$ changes, the shape of  the average time above threshold is fairly similar. In all cases the average excess time decreases monotonically with the threshold level, with a fast drop for small threshold values. This is followed by a slow tapering off for large threshold values. For the range of intermittency parameters considered here, the average excess time is of the order of the pulse duration or shorter for large threshold values. The average time above threshold decreases by about half a decade for each tenfold increase in $\gamma$, but the functional shape varies little. Indeed, it can be shown that for given $\gamma$ and $\lambda$, $\avT /\td$ scales as $\mSn / \Sn$ in the limit $\Sn \to \infty$.
As the threshold value increases above the mean signal value, up-crossings of the threshold become fewer while the signal spends less time in total above the threshold. Evidently these two effects nearly cancel, and the average excess time decreases slowly with increasing threshold level.

Considering comparisons to experimental data, the results presented here provide two major improvements over the classical Rice's formula in the case of intermittent fluctuations. Firstly, any discrepancy between the normal limit for excess time statistics and measurement data has previously been interpreted as a signature of intermittency in the process. The formulas derived here quantifies the level of intermittency by the model parameters $\lambda$ and $\gamma$. Secondly, Rice's formula requires the rms-value of the derivative of the signal, which is difficult if not impossible to estimate for discretely sampled data. In contrast, estimates for $\lambda$ and $\gamma$ can be found from the signal using the lowest order moments of $\Sn$ and its correlation function. \cite{theodorsen,kube} 


In conclusion, a previously suggested stochastic model for intermittent fluctuations in the scrape-off layer of magnetically confined plasmas has been considered. The model consists of a super-position of pulses with a fixed, exponential pulse shape and exponentially distributed amplitudes arriving according to a Poisson process. In this contribution, the joint probability density function of the random variable and its derivative is derived and is used to obtain predictions for level crossings and average excess times for fluctuations above a given threshold. These predictions depend on two model parameters, the intermittency parameter $\gamma$ and the pulse shape asymmetry parameter $\lambda$. It is shown that the functional shape of the rate of level crossings with the threshold level is strongly dependent on the intermittency parameter $\gamma$ of the process, while the functional shape of the average excess time varies little with the parameter $\gamma$, suggesting that the rate of level crossings might be a more useful tool in comparing the model to experimental data in order to assess intermittency effects. In both cases, the functional shape does not depend on $\lambda$.

Even though the total time above a given threshold level may be the same for realizations of two different intermittent processes, this can be realized through either many short plasma bursts or few but long lasting bursts events. This may have profound implications for plasma-wall interactions in magnetically confined plasmas, since long lasting, large amplitude events can lead to severe damaging while the system can recover from the damaging impacts of shorter burst events depending on their frequency of occurrence.\cite{sato,fattorini} Thus accurately predicting the rate of level crossings and average excess times for an intermittent process is of considerable interest to statistical modelling of fluctuations in the boundary region of magnetically confined plasmas. In future work, the novel predictions presented here will be compared to experimental measurement data from the scrape-off layer of magnetically confined plasmas.


\section*{Acknowledgements}

This work was supported with financial subvention from the Research Council of Norway under grant 240510/F20. Discussions with H. ~L.~P{\'e}cseli are gratefully acknowledged.



\begin{thebibliography}{99}
%
%
\bibitem{pitts}
R.~A.~Pitts, \PPCF\ {\bf 47}, B303 (2005).
%
\bibitem{lipschultz}
B.~Lipschultz, X.~Bonnin, G.~Counsell \etal, \NF\ {\bf 47}, 1189 (2007).
%
\bibitem{dmz}
D.~A.~D'{I}ppolito, J.~R.~Myra and S.~J.~Zweben, \PP\ {\bf 18}, 060501 (2011).
%
%
\bibitem{garcia-acm}
O.~E.~Garcia, S.~M.~Fritzner, R.~Kube, I.~Cziegler, B.~LaBombard and J.~L.~Terry, \PP\ {\bf 20}, 055901 (2013); \JNM\ {\bf 438}, S180 (2013).
%
\bibitem{garcia-tcv}
O.~E.~Garcia, J.~Horacek and R.~A.~Pitts, \NF\ {\bf 55}, 062002 (2015).
%
\bibitem{theodorsen}
  A.~Theodorsen, O.~E.~Garcia, J.~Horacek, R.~Kube and R.~A.~Pitts, \PPCF\ {\bf 58} 044006 (2016).
%
\bibitem{kube}
  R.~Kube, A.~Theodorsen, O.~E.~Garcia, B.~La{B}ombard and J.~L.~Terry, \PPCF\ {\bf 58}, 054001 (2016).
%
%
\bibitem{rice}
S.~O.~Rice, \BTSJ\ {\bf 23}, 282 (1944).
%
\bibitem{garcia-shotnoise}
  O.~E.~Garcia, \PRL\ {\bf 108}, 265001 (2012); O.~E.~Garcia, A.~Theodorsen, R.~Kube and H.~L.~P{\'e}cseli, submitted to \PP.
%
\bibitem{kg}
R.~Kube and O.~E.~Garcia, \PP\ {\bf 22}, 012502 (2015).
%
\bibitem{sattin}
  F.~Sattin, M.~Agostini, P.~Scarin, N.~Vianello, R.~Cavazzana, L.~Marrelli, G.~Serianni, S.~J.~Zweben, R.~J.~Maqueda, Y.~Yagi, H.~Sakakita, H.~Koguchi, S.~Kiyama, Y.~Hirano and J.~L.~Terry, \PPCF\ {\bf 51}, 055013 (2009).
%
\bibitem{bergsaker}
  A.~S.~Bergsaker, {\AA}.~Fredriksen, H.~L.~P{\'e}cseli and J.~K.~Trulsen, \PS\ {\bf 90}, 108005 (2015).
%
%

%
%
%
%
%
%
\bibitem{rice-level}
S.~O.~Rice, \BTSJ\ {\bf 24}, 46 (1945).
%
\bibitem{kristensen}
L.~Kristensen, M.~Casanova, M.~S.~Courtney and I.~Troen, \BLM\ {\bf 55}, 91 (1991).
%
\bibitem{sato}
H.~Sato, H.~L.~P{\'e}cseli, and J.~Trulsen, \JGR\ {\bf 117}, A03329 (2012).
%
\bibitem{fattorini}
  L.~Fattorini, {\AA}.~Fredriksen, H.~L.~P{\'e}cseli, C.~Riccardi and J.~K.~Trulsen,  \PPCF\ {\bf 54}, 085017 (2012).
%
\bibitem{dalmao}
F.~Dalmao and E.~Mordecki, \textit{Extremes} {\bf 18}, 15 (2015).
%
\end{thebibliography}

\end{document}